# Breakdown of agreement between non-relativistic and relativistic quantum dynamical predictions for the periodically kicked rotor

Boon Leong Lan[*] and Rui Jian Chu
*Electrical and Computer Systems Engineering & Advanced Engineering Platform, School of Engineering, Monash University, 47500 Bandar Sunway, Malaysi*a
[*]Corresponding author. Email: lan.boon.leong@monash.edu

**ABSTRACT**

As an extension of our recent work, we show, both theoretically and numerically, that the breakdown of agreement between the non-relativistic and relativistic quantum dynamical predictions in the non-relativistic regime also occurs for an electron subjected to a time-dependent force.

To study the dynamics of quantum systems in the non-relativistic regime, the standard practice is to use non-relativistic quantum theory, instead of the more complicated relativistic version. This is because it is conventionally thought the former will always yield similar results to the latter [1]. However, we recently proved that this expectation is not true in general for the motion of an electron subjected to a time-independent potential [2]. Furthermore, our numerical results for the free rotor and hydrogen atom showed the agreement between the two theories can break down quickly and the different predictions could be tested experimentally for hydrogen atom [2]. Here we extend our work to the case of a time-dependent potential, in particular, to the periodically kicked rotor.

Consider an electron (with charge $e$ and rest mass $m_0$) that is constrained to move in a circle of radius $R$ and subjected to an angular-position-dependent electric potential $V\cos(\emptyset)$ that is turned on periodically, with period $T$, for an instant. In the nonrelativistic case, if the wave function is expanded in the non-relativistic energy eigenstates of the free rotor

$$\psi_{NR}(\emptyset, t) = \sum_{q=-\infty}^{\infty} a_q(t) \frac{1}{\sqrt{2\pi}} e^{iq\emptyset} \qquad [1]$$

with energies $\varepsilon_q = \frac{\hbar^2 q^2}{2m_0 R^2}$, the solution of the time-dependent Schrodinger equation leads to an exact mapping of the expansion coefficients from just before the n*th* kick to just before the (n+1)*th* kick

$$a_r((n+1)T^-) = \sum_{q=-\infty}^{\infty} a_q(nT^-) i^{r-q} J_{r-q}(\kappa) e^{-\frac{i\tau r^2}{2}}, \qquad [2]$$

which depends on two dimensionless parameters $\kappa \equiv \frac{eVT}{\hbar}$ and $\tau \equiv \frac{\hbar T}{m_0 R^2}$ that are related to the properties of the kicking force. Details of the derivation of the non-relativistic map Eq. (2) are given in [3].

In the relativistic case, the wave function is expanded in the relativistic energy eigenstates of the free rotor given in [2]

$$\psi(\emptyset, t) = \sum_{q=-\infty}^{\infty} A_q(t) \begin{bmatrix} 1 \\ 0 \\ 0 \\ i\sqrt{\frac{1}{2\pi N_q^2} - 1} \end{bmatrix} N_q e^{iq\emptyset}, \qquad [3]$$

with energies $E_q = m_0 c^2 \left[\sqrt{1 + \gamma^2 q^2} - 1\right]$, where

$$N_q = \frac{1}{\sqrt{2\pi \left[1 + \frac{\gamma^2 q^2}{\left(\sqrt{1+\gamma^2 q^2} + 1\right)^2}\right]}}, \qquad [4]$$

is the normalization constant, and $\gamma \equiv \frac{\hbar}{m_0 c R}$. The solution of the time-dependent Dirac equation also leads to an exact mapping of the expansion coefficients from just before the n*th* kick to just before the (n+1)*th* kick

$$A_r((n+1)T^-) = \sum_{q=-\infty}^{\infty} A_q(nT^-) i^{r-q} J_{r-q}(\kappa) \Omega_{r,q} e^{-i\tau\gamma^{-2}\left(\sqrt{1+\gamma^2 r^2}-1\right)}, \qquad [5]$$

which also depends on the dimensionless parameters $\kappa$ and $\tau$, where *J* is a Bessel function of the first kind with integer order, and

$$\Omega_{r,q} = 2\pi N_r N_q + \sqrt{(1 - 2\pi N_r^2)(1 - 2\pi N_q^2)}. \qquad [6]$$

The method of deriving the relativistic map Eq. (5) is the same as the derivation of the non-relativistic map Eq. (2).

In the non-relativistic regime, i.e., $\gamma q \ll 1$ [4], the relativistic wave function is given by

$$\psi(\emptyset, t) \approx \sum_{q=-\infty}^{\infty} A_q(t) \begin{bmatrix} 1 \\ 0 \\ 0 \\ 0 \end{bmatrix} \frac{1}{\sqrt{2\pi}} e^{iq\emptyset}, \qquad [7]$$

since $N_q \approx \frac{1}{\sqrt{2\pi}}$. Keeping the first three terms in the binomial expansion of $\sqrt{1+\gamma^2 r^2}$ yields

$$A_r((n+1)T^-) \approx \left(\sum_{q=-\infty}^{\infty} A_q(nT^-) i^{r-q} J_{r-q}(\kappa) e^{-\frac{i\tau r^2}{2}}\right) e^{\frac{i\tau\gamma^2 r^4}{8}}, \qquad [8]$$

for the relativistic map since $\Omega_{r,q} \approx 1$. In comparison to the non-relativistic map, the relativistic map has an extra phase factor $e^{\frac{i\tau\gamma^2 r^4}{8}} = e^{-\frac{i\delta_r T}{\hbar}}$, where $\delta_r = -\frac{\hbar^4 r^4}{8 m_0^3 c^2 R^4}$ is the leading term in the difference between the relativistic and non-relativistic energies for the free rotor. This extra phase factor will cause the relativistic and non-relativistic time-dependent wave functions, with the same initial expansion coefficients, to diverge and eventually disagree. If $\tau \ll \gamma^2$ (which implies the kicking period $T$ is $\ll \frac{\hbar}{m_0 c^2}$), the extra phase factor $e^{\frac{i\tau\gamma^2 r^4}{8}}$ is $\approx 1$ since $\frac{\tau\gamma^2 r^4}{8} \ll 1$. In this case, the breakdown of agreement between the relativistic and non-relativistic wave functions will take a long time to occur. However, if the condition $\tau \ll \gamma^2$ is not satisfied, the breakdown of agreement will be faster.

An example of the breakdown of agreement is given in Fig. 1, where $\gamma = 2 \times 10^{-7}$, $\kappa = 10^{-7}$ and $\tau = 10^6$. The relativistic and non-relativistic initial expansion coefficients are chosen to be

$$A_n(0) = a_n(0) = \left(\frac{2\sigma_0^2}{\pi}\right)^{1/4} exp(-in\theta_0) \, exp(-\sigma_0^2(n-\bar{n})^2), \qquad [9]$$

where $\sigma_0 = 0.8$, $\theta_0 = \pi$ and $\bar{n} = 48$. For $\sigma_0 < 1$, the relativistic and non-relativistic initial probability densities are approximately Gaussians in the angle interval $[0, 2\pi]$ with mean $\theta_0$ and standard deviation $\sigma_0$. For both the mean (Fig. 1(a)) and standard deviation (Fig. 1(b)), the relative difference between the relativistic and non-relativistic values fluctuates as it grows with the kick number. For the mean (standard deviation), the relative difference is 5% (3.7%) and 50% (19%) at kick 52 and 509, respectively. The relativistic and non-relativistic probability densities are still close at kick 52 (Fig. 1(c)) but they are very different at kick 509 (Fig. 1(d)).

**References**

[1] Messiah, *Quantum Mechanics, Vol. II* (North-Holland Publishing, Amsterdam, 1961), Chap. XX.
[2] B. L. Lan, M. Pourzand and R. J. Chu, Breakdown of agreement between non-relativistic and relativistic quantum dynamical predictions in the non-relativistic regime, Results in Physics 12, 147-152 (2019).

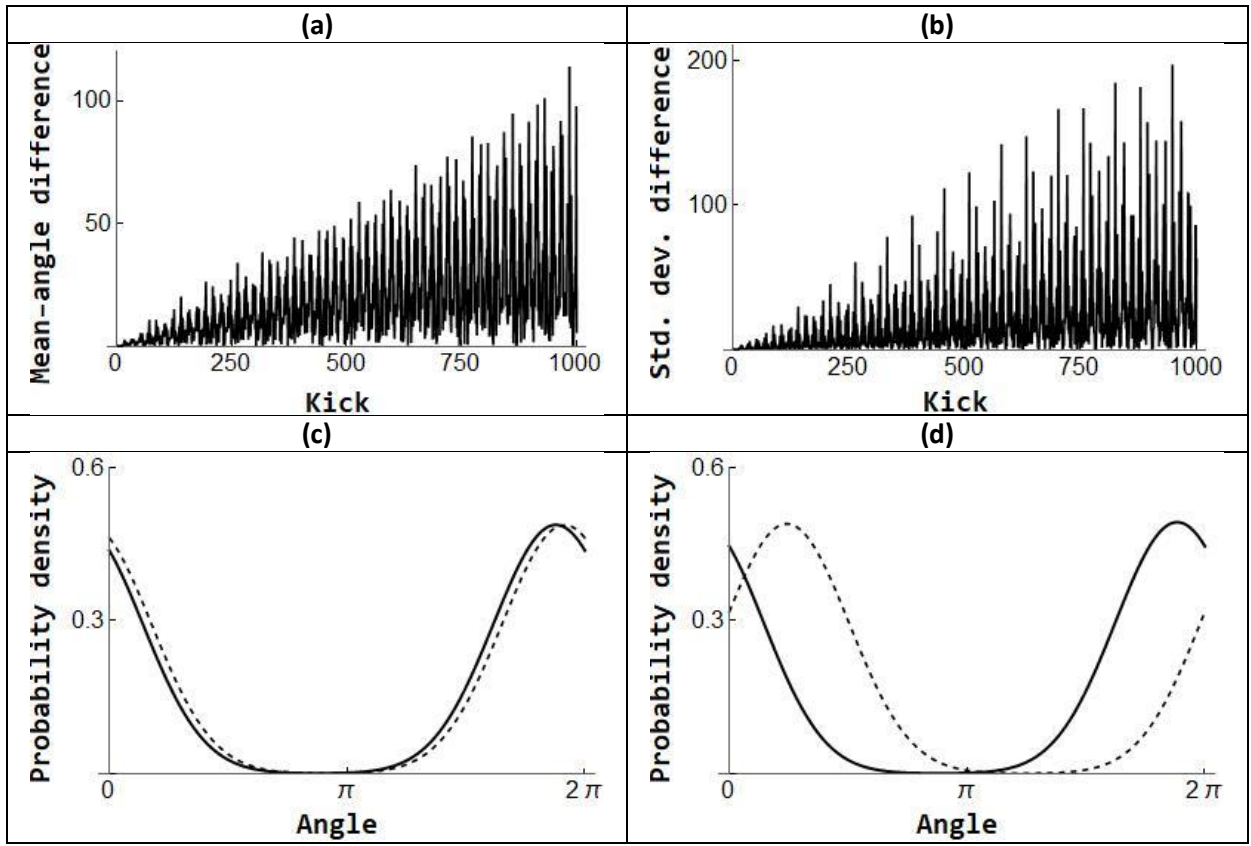

**Figure 1.** Magnitude of the relative difference (calculated as a percentage with respect to the relativistic value) between the relativistic and non-relativistic values for the (a) mean angle, (b) angle standard deviation. Relativistic (solid line) and non-relativistic (dash line) probability densities at (c) kick 52, (d) kick 509.